\documentclass[12pt]{iopart}
\usepackage{graphicx}
\usepackage{url}
\usepackage{xcolor}

\usepackage{iopams}  
\begin{document}

\title[SurferBot]{SurferBot: A wave-propelled aquatic vibrobot}

\author{Eugene Rhee$^1$, Robert Hunt$^1$, Stuart J. Thomson$^{1,2}$, and Daniel M. Harris$^1$}
\address{$^1$ Brown University, Center for Fluid Mechanics and School of Engineering, 184 Hope St., Providence, Rhode Island 02912, USA}
\address{$^2$ University of Bristol, Department of Engineering Mathematics, Ada Lovelace Building,
University Walk, Bristol, BS8 1TW, UK}
\ead{daniel\_harris3@brown.edu}
\vspace{10pt}
\begin{indented}
\item[]May 2022
\end{indented}

\begin{abstract}

Nature has evolved a vast array of strategies for propulsion at the air-fluid interface. Inspired by a survival mechanism initiated by the honeybee (\emph{Apis mellifera}) trapped on the surface of water, we here present the \emph{SurferBot}: a centimeter-scale vibrating robotic device that self-propels on a fluid surface using analogous hydrodynamic mechanisms as the stricken honeybee. This low-cost and easily assembled device is capable of rectilinear motion thanks to forces arising from a wave-generated, unbalanced momentum flux, achieving speeds on the order of centimeters per second. Owing to the dimensions of the SurferBot and amplitude of the capillary wave field, we find that the magnitude of the propulsive force is similar to that of the honeybee. In addition to a detailed description of the fluid mechanics underpinning the SurferBot propulsion, other modes of SurferBot locomotion are discussed. More broadly, we propose that the SurferBot can be used to explore fundamental aspects of active and driven particles at fluid interfaces, as well as in robotics and fluid mechanics pedagogy. 
\end{abstract}

%
%Uncomment for keywords
\vspace{2pc}
\noindent{\it Keywords}: Robotics, capillarity, honeybee, waves, self-propulsion, frugal science, pedagogy
%\maketitle

\section{Introduction}

Biological organisms have evolved myriad strategies for propulsion at the air-water interface, with the relevant fluid physics generally depending on length scale \cite{bush2006walking}.  For organisms at the millimeter and centimeter scale, surface tension effects are frequently exploited for both flotation and propulsion \cite{vella2015floating}. There are two principal mechanisms by which these capillary-scale organisms may propel on a flat fluid interface: chemically driven propulsion by which a surface tension gradient, or Marangoni stress, is generated; or mechanical propulsion wherein kinematic motion of the organism results in a net hydrodynamic thrust.  Techniques that exploit mechanical propulsion rely on a transfer of momentum to the fluid through reciprocal kinematic motions of its appendages, a phenomenon exemplified by water striders \cite{hu2003hydrodynamics}, whirligig beetles \cite{voise2010management}, and water boatmen \cite{ngo2014hydrodynamics}, among others.  This momentum flux generally takes the form of propagating waves or coherent subsurface flows, such as vortices \cite{buhler2007impulsive,gao2011numerical,steinmann2018unsteady}.

This vast array of surface propulsion strategies identified in nature has inspired the development of a similarly broad range of biomimetic robotic devices, with varying levels of complexity and control \cite{hu2010water, yuan2012bio, burton2013biomimicry, ozcan2014stride, koh2015jumping, philamore2015row, chen2018controllable, grosjean2018surface, feldmann2021can, timm2021remotely}.  Such small-scale autonomous robotic devices have been proposed for applications such as environmental monitoring, surveillance, and pedagogical purposes.  

%Due to reduction in wetted area, operation at the air-water interface could be more efficient some case as a result of reduced drag forces \cite{chen2018controllable}.

In the present work, our attention is drawn to the honeybee (\emph{Apis mellifera}), whose wings can become trapped at an air-water interface due to surface tension. This unfortunate event can occur when the honeybee hovers too close to, and accidentally brushes, the water surface. It was recently demonstrated that the stricken honeybee initiates a survival mechanism by oscillating its wings to generate fore-aft asymmetric capillary waves and fluid flow, propelling itself forward in an attempt to reach safety \cite{roh2019honeybees}.  Inspired by this natural observation, we present the design and characterization of a simple, low-cost, untethered, robotic device that propels itself forward using analogous hydrodynamic mechanisms as the surface-bound honeybee.  Our device, referred to herein as \emph{SurferBot}, consists of a centimeter-scale thin, flat plate that is pinned at the air-water interface and driven into oscillation by an on-board vibration motor.  The vibration motor is positioned along the plate so as to generate an asymmetric rocking motion, which in turn results in an asymmetric capillary wavefield and concomitant fluid flow.  These mechanisms lead to steady unidirectional propulsion with speeds on the order of centimeters per second.  Direct measurements of the global wavefield and surface fluid flow are presented, from which the propulsive forces on the SurferBot are estimated. Owing to the dimensions of the SurferBot and the frequency and amplitude of the capillary wave field, we find that the propulsive forces are very similar in magnitude to those generated by the honeybee. 

This paper is organized as follows. In \S\ref{sec:methodsAll}, we provide an account of the methods used in the design and manufacturing of the SurferBot, as well as the techniques employed to measure both the capillary wavefield and accompanying fluid flows resulting from the vibration of the SurferBot. Experimental results of both the SurferBot and fluid motion (acquired using the techniques outlined in \S\ref{sec:methodsAll}) are presented in \S\ref{sec:resultsAll}, along with a comparison of the propulsive force generated by the SurferBot to that generated by the honeybee. A summary of our primary results and comparison to prior work is presented in \S\ref{sec:discussion}, with concluding remarks and suggestions for future work discussed in \S\ref{sec:conclusion}.

%Due to the geometry of the plate, the motion is principally resisted by skin friction along the base of the SurferBot.

% Bees \cite{roh2019honeybees}.
% Water walking review \cite{bush2006walking}.
% Water strider Robostrider \cite{hu2003hydrodynamics}.
% Water walking devices \cite{hu2010water,yuan2012bio}.
% Electrowetting wave propulsion \cite{chung2009electrowetting}.
% Marangoni surfer \cite{timm2021remotely}?
% Row-bot \cite{philamore2015row}.

%This should be concise and describe the nature of the problem under investigation and its background. It should also set your work in the context of previous research, citing relevant references. Introductions should expand on highly specialised terms and abbreviations used in the article to make it accessible for readers.

\section{Methods}
\label{sec:methodsAll}
The SurferBot is a simple, low-cost device that achieves self-propulsion by the action of a vibrating motor mounted on the SurferBot's back. We begin with an account of the SurferBot design and construction, followed by details of the tracking procedure used to record the trajectory of the device. This section concludes with a summary of the methods used to visualize and reconstruct both the capillary wave field radiating outward from the periphery of the SurferBot and the accompanying surface fluid flows. 

\subsection{SurferBot design and manufacturing}\label{3d}

%The surfer was fabricated through 3-D printing on a Elegoo Mars 2 Pro resin printer using Anycubic clear UV sensitive resin. A magnetic printer building plate with a magnetic sheet was used, which allowed for easy release and maintained a smooth bottom surface of the surfer. Due to the size constraint of the build plate, 3 surfers were able to be printed at a time. Finished surfer prints were submerged and shaken by hand in a bath of IPA, and rinsed off using deionized water. The surfer was modeled with Fusion 360, with a rectangular body that is 50 mm long and 30 mm wide. The corners were rounded with a 2 mm radius, in order to maintain a good contact line and avoid pinning. Four structural support systems were also added to retain a straight and flat surfer body. 

%Eccentric mass motors, 2 mm thick with a diameter of 10 mm, were used as the vibration method. The motor wires were stripped until 2 cm from the motor, and attention was taken to avoid short circuiting the battery and maintain good electrical contact with the coin cell battery. Plastic tweezers, instead of metal, were mainly used to push the motor wires to press fit with the battery, also in order to avoid short circuiting the battery. 

\begin{figure}
\centering
\includegraphics{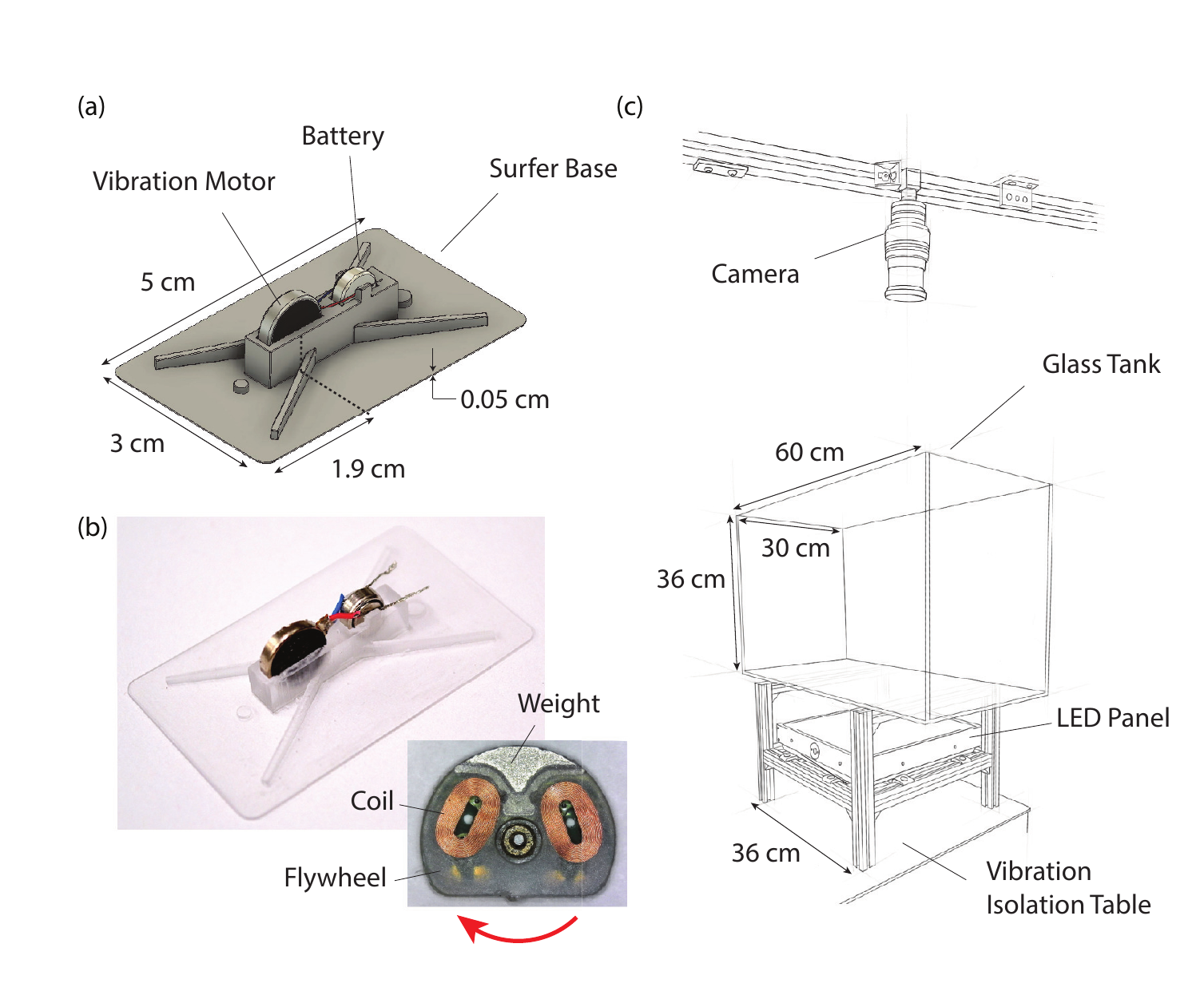}
\caption{(a) CAD rendering of an assembled SurferBot. The position of the motor and battery housing relative to the center of the base creates a desired asymmetry along the SurferBot's length. (b) Assembled SurferBot used in experiments. Inset: internal view of the vibration motor's eccentric mass. (c) Schematic of experimental setup.}
\label{fig:schem}
\end{figure}

%Figure 1: schematic and aesthetic picture of assembled SurferBot. dimensions labeled somewhere.
The SurferBot consists of three primary components: a 3D printed rectangular base; a vibration motor containing an eccentric mass mounted on a rotating flywheel; and an alkaline button cell battery (Figures \ref{fig:schem}(a) and (b)). The SurferBot base, vibration motor, and battery weighed approximately 1.6, 0.7, and 0.3 grams respectively, for a total mass of 2.6 grams. The total cost of materials for a single SurferBot is less than \$1. We note that the manufacturing techniques presented here are employed for scientific accuracy and repeatability.  However, as discussed in \S\ref{sec:discussion}, an even simpler version of the SurferBot can be readily realized for pedagogical purposes.

\vspace{0.1in}

\noindent \emph{Base} -- The SurferBot base was modeled using Autodesk Fusion 360 software and then 3D printed using an Elegoo Mars 2 Pro Stereolithography printer with Anycubic clear photopolymer resin. CAD files of the SurferBot are included as part of the Supplementary Material. A magnetic sheet and removable build plate (Biqu WZC000408) were installed on the printer bed that allowed for the SurferBot base to be printed directly on the printer bed without supports and then easily released. Once printed, the SurferBot base was submerged and shaken by hand in a bath of isopropyl alcohol and then rinsed using deionized water to remove any residual uncured resin. The dimensions of the thin rectangular base were 5 cm $\times$ 3 cm $\times$ 0.05 cm. {\color{black} The SurferBot is supported at the interface due to the combined effects of surface tension along the perimeter and hydrostatic pressure across the base area \cite{pan2009miniature}.  The base dimensions were chosen to be sufficiently large to support the mass of the 3D-printed structure and electronics (2.6 grams), although it was observed that an additional 1.3 grams could be supported before the device sank.}   The thickness of the SurferBot base was sensitive to alignment of the printing bed but variations were typically 150 $\mu$m or less.  The corners of the base were rounded with a 2 mm radius in order to maintain the contact line along the perimeter of the base. A support structure consisting of four arms extending radially from the motor and battery housing was added to increase rigidity and reduce warping of the base. An essential component in the SurferBot's design is the front-rear asymmetry of the device created by the position of the motor and battery housing, which gives rise to a fore-aft asymmetric capillary wavefield and steady unidirectional propulsion. {\color{black} In particular, the motor is centered 6 mm behind the geometric center of the base and the overall center of mass located 1 mm behind the center.}  Note that the mass is intended to be evenly distributed along the \emph{width} of the SurferBot to inhibit rolling about the long axis of the device.

\vspace{0.1in}

\noindent\emph{Electronics} -- Commercially available vibration motors (BestTong A00000117, 10mm OD, 2mm thick), powered by a coin cell battery (TRUSTYIWEN 377, 6.7 mm OD, 2.6 mm thick), were used to apply oscillatory forcing. To best ensure uniformity between trials, batteries with voltages between 1.57 to 1.59 V were selected for our experiments. The motors exhibited great variability in performance, however steps were taken to ensure similarly performing motors were selected for consistency between experimental trials (see \S\ref{methodmotor}). The motor wires were precisely stripped 2 mm from the motor body to avoid short circuiting the battery. A support post next to the battery with a rounded edge allowed for the motor wires to be press fit into place and maintain electrical contact with the battery. Plastic tweezers were used to push the motor wires into place at the start of the experiment.  Once connected, the battery continuously drives the motor for approximately 12 minutes.

\subsection{Vibration motor characterization}\label{methodmotor}

\begin{figure}
\centering
\includegraphics{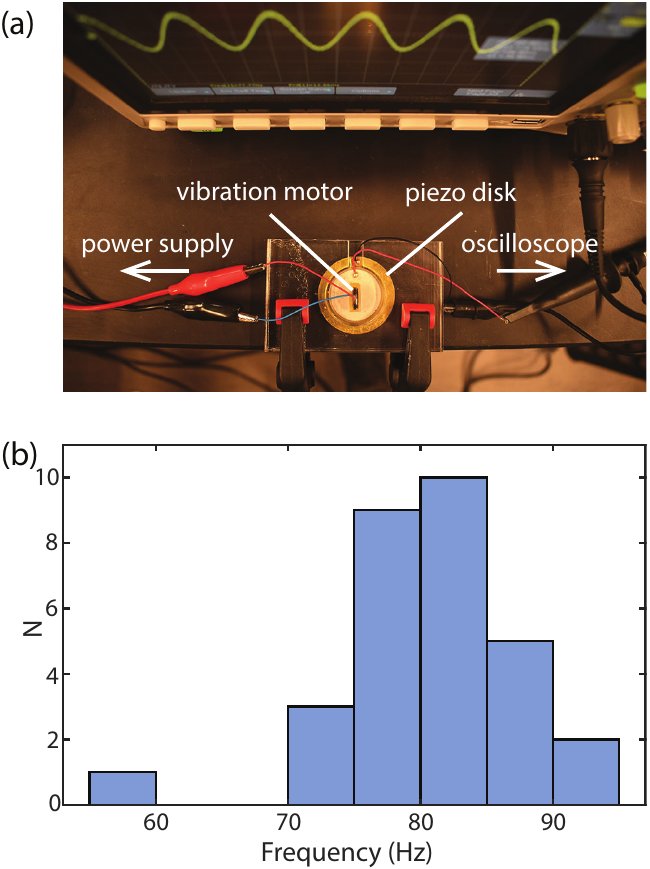}
\caption{(a) Picture of vibration motor characterization setup from above.  A piezoelectric disk is used to measure the frequency and relative vibration amplitude of a motor. (b) Histogram of motor frequencies from a pack of 30 motors, measured from piezoelectric disk output on an oscilloscope. Motors with initial frequency between 80 and 82 Hz and amplitude between 1.20 and 1.24 V were chosen for all experiments.}
\label{fig:motorchar}
\end{figure}

To characterize the performance of the vibration motors, a 3.5 cm diameter piezoelectric disk (CUI Devices CEB-35D26) was clamped along its perimeter between two sheets of laser-cut acrylic (0.25 inches thick), which were welded together at one end (Figure \ref{fig:motorchar}(a)). The motors were held fixed in a 3D printed housing derived from the full SurferBot CAD model to replicate operating constraints, and this housing was attached to the piezoelectric disk using double-sided tape. The motors were connected to a switching DC power supply (Tekpower TP3005P) set at constant voltage of 1.0 V. {\color{black} Although the open-circuit voltage of the battery is nominally 1.5 V, 1.0 V is chosen for these tests to be consistent with the observed voltage across the motor during operation when actively powered by the coin-cell battery.}  The nearly sinusoidal output from the piezoelectric disk was measured using an oscilloscope (Siglent SDS-1202X-E). The measured frequency and peak-to-peak amplitude of motors typically varied between 70 and 90 Hz and 1.0 and 1.5 V, respectively. A sample of the frequency distribution from a typical pack of 30 motors is provided in Figure \ref{fig:motorchar}(b). Motors with frequency between 80 and 82 Hz and amplitudes between 1.20 and 1.24 V were selected for use in all experiments. Despite these initial selection criteria, it was observed that further changes occur in the frequency and amplitude during the lifetime of the motor, with no evident pattern.

%Figure 2: Labeled photo of Robert's piezo characterization setup with oscilloscope waveform in image. Histogram(s) of performance. (single column)

% \subsection{Testing facility}
% Eugene.
% Figure 3: Eugene sketch (with labels and dimensions). (single column).

\subsection{Planar position tracking}\label{sec:methodposition}

In order to visualize the trajectory of the SurferBot over an appreciable distance, experiments were performed in a large glass tank filled to a depth of 5 mm with deionized water and leveled to ensure uniformity of depth. A schematic of the experimental setup and dimensions is shown in Figure \ref{fig:schem}(c). The tank was supported from below by an extruded aluminum frame bolted to an optical table. A 1 megapixel camera (Allied Vision Mako U-130B), fitted with a Nikon AF Nikkor 24-120mm lens, was placed overhead, supported by an extruded aluminum rail 96 cm above the water surface. A 30 cm square LED panel (Dracast LED1000 Pro) was placed 9 cm below the bottom surface of the tank and supported by the aluminum frame. Diffuser paper was placed on the LED panel and the outer bottom surface of the tank to improve background light uniformity. 

Before each set of experimental trials and prior to filling the tank with water, the tank was rinsed thoroughly with deionized water. Once the tank was filled, a SurferBot was assembled and placed on the water surface with tweezers. By submerging a thin wooden rod just in front of the SurferBot's front edge, the SurferBot was first restrained at one end of the tank to allow for transient surface disturbances to decay. The SurferBot was then released by lifting the rod and its trajectory was recorded at 30 frames-per-second (fps) using \texttt{imaqtool} in MATLAB.  The top surfaces of two small extruded cylinders (see Figure \ref{fig:schem}(a)) on the front and back of the SurferBot were dyed with marker ink. This allowed the position of the SurferBot to be obtained during post-processing of the experimental trajectories using the circular Hough transform in MATLAB. The video frames were binarized and cropped using a 20 pixel sliding interrogation window. The first four seconds of each experimental trial were excluded when calculating the average velocity magnitude to avoid periods of transient acceleration.

\subsection{Sideview kinematics}\label{methodsideview}
The SurferBot kinematics were likewise studied from the side using a 1 megapixel high-speed camera (Vision Research VEO 1310L) recording at 5000 fps. This data was captured in a smaller glass tank (9.5 cm width, 29.5 cm depth, 9 cm height), filled to 5 cm depth for ease of visualization and backlit using the square LED panel. A Nikon AF Micro Nikkor 200 mm lens was coupled with a Kenko 2x Teleplus Pro 300 teleconverter for a real-space pixel extent of 65 microns. The position of the SurferBot lower base along the front and rear edges was registered to subpixel precision using the MATLAB function \texttt{imregtform} with \texttt{transformType} set to \texttt{translation} and an interrogation window size of $200\times 200$ pixels. Front and back edges of a single SurferBot were captured independently as the SurferBot passed through the field of view, their paths synchronized by phase matching their horizontal motion.

\subsection{Wavefield reconstruction and visualization}\label{fcd}

The propagating capillary wavefield generated by the vibrating SurferBot is reconstructed using the Fast Checkerboard Demodulation (FCD) method \cite{wildeman2018real}.  The FCD method is a recent variation to the widely used synthetic Schlieren imaging (SSI) technique for reconstructing surface waves on a transparent fluid \cite{moisy2009synthetic}.  One significant attraction of SSI methods is that they are relatively accessible, with a camera and printer representing the only hardware requirements. In general, SSI techniques extract a pixel displacement field, $\mathbf{u}(x,y)$, by comparing an image of a reference background pattern to one optically distorted by deformations of a fluid interface. If the camera is placed sufficiently far above the free surface, to leading order, the pixel displacement field is proportional to the gradient of the surface height profile, $h(x,y)$, \emph{via}
\begin{equation}
    \mathbf{u}(x,y) = -H\left(1-\frac{n_{a}}{n_{w}}\right) \nabla h(x,y), \label{eqn:SSI}
\end{equation}
where $H$ is the distance between the undisturbed interface and the background pattern and $n_a/n_w=0.75$ is the ratio of the indices of refraction of air and water \cite{wildeman2018real, moisy2009synthetic}.  In order to estimate $h(x,y)$, one must invert the gradient field, $\nabla h(x,y)$.  Inversion of $\nabla h(x,y)$ is generally accomplished by numerically solving an overdetermined set of linear equations using least-squares.

The FCD method uses a periodic checkerboard pattern as the background image and determines $\mathbf{u}(x,y)$ using spatial Fourier demodulation \cite{grediac2016grid}. The FCD method has been shown to be at least as accurate as more traditional SSI methods that rely on a random dot pattern and Digital Image Correlation (DIC) algorithm, but is significantly faster. The MATLAB source code for the FCD method is publicly available on GitHub \cite{wildemangithub}.

The overhead imaging configuration shown in Figure \ref{fig:schem}(c) is used to record the SurferBot and background pattern at 30 fps with a spatial resolution of 5.9 pixels/mm.  A square checkerboard pattern with periodicity of 1.16 mm was printed on a clear transparency with a resolution of 600 dots-per-inch and secured to the bottom of the tank just beneath the fluid. Hence, $H$ coincides with the depth of the fluid. The checkerboard pattern corresponds to approximately 7 pixels for each wavelength, consistent with practical guidelines suggested in \cite{wildeman2018real}.

Per Equation (\ref{eqn:SSI}), $\mathbf{u}(x,y)$ is directly proportional to $H$. As discussed in \S\ref{sec:methodposition}, the depth of the working fluid was set at a relatively shallow $H=5.0$ mm which maximizes resolution while avoiding large pixel displacements (known to cause SSI methods to become inaccurate). In a small region surrounding the SurferBot (extending about two capillary lengths, or $\sim5$ mm), some artifacts due to high surface distortion and the mask nevertheless appear in the data. Outside of this narrow border surrounding the SurferBot, all criteria for faithful surface reconstruction using the FCD method were determined to be satisfied and the details of the mask had no influence on the reconstructed wave profile.  

For processing the images, the SurferBot position was first identified using the tracking dots (\S\ref{sec:methodposition}) and the images were then masked using a $5$ cm $\times$ $3$ cm rectangular region to ensure that the region occupied by the SurferBot was not included in the wavefield reconstruction algorithm. The displacement field $\mathbf{u}(x,y)$ was determined using the MATLAB-based FCD code provided at \cite{wildemangithub} and subsequently $\nabla h(x,y)$ was inverted using the \texttt{intgrad2} function in MATLAB \cite{d2009inverse}.  Finally, the surface height profile $h(x,y)$ was processed using a spatial band-pass filter, primarily to remove any nonphysical large wavelength distortions predicted by the method as described in \cite{damiano2016surface}.

\begin{figure}
\centering
\includegraphics{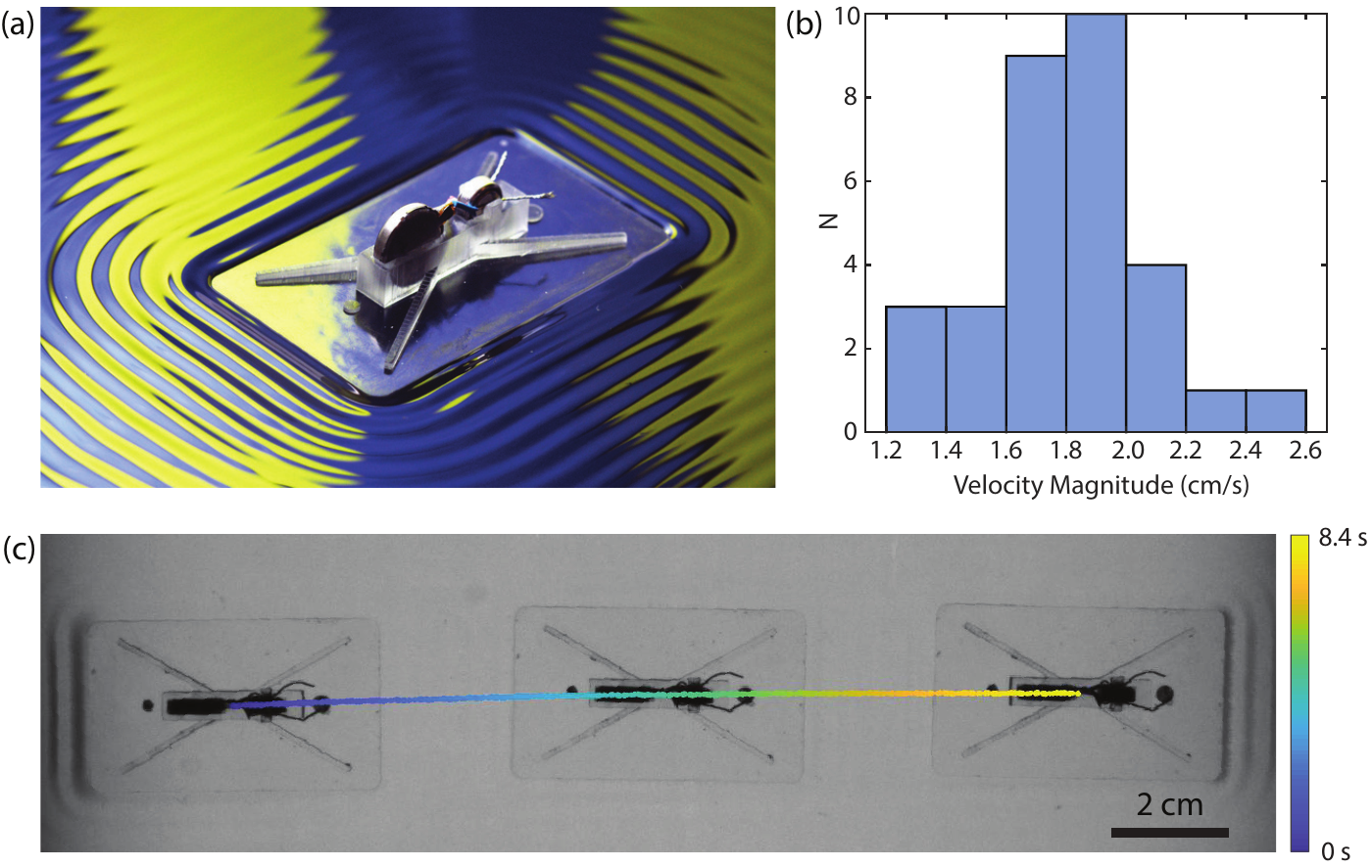}
\caption{(a) Photograph of a SurferBot during operation. The reflected color emphasizes the deformation of the fluid surface due to the SurferBot vibration. (b) Histogram of average SurferBot velocity magnitudes for each trial with tracking and analysis methods described in \S\ref{sec:methodposition}.  Data was collected with six independent SurferBots.  (c) Overhead view of SurferBot motion (left-to-right) pictured at three separate times. The colored dots represent the tracked position of the center over 8.4 seconds. See Supplementary Materials for related videos.}
\label{speeds}
\end{figure}

In complement to the quantitative wavefield reconstruction technique described above, qualitative images and videos of the surface waves are also taken by visualizing the reflection of an illuminated color pattern distorted by the wavy surface \cite{harris2017visualization}, an example of which is shown in Figure \ref{speeds}(a).

\subsection{Surface flow characterization}\label{methodsurface}

The vibration of the SurferBot establishes a surface flow field accompanying the radiating capillary wavefield. To visualize the flow field, buoyant hollow glass microspheres of diameter approximately 70 $\mu$m (Fibreglast Part \#22) were suspended at the air-water interface and used as tracer particles \cite{colombi2021three}. To avoid clumping, the particles were stirred before each experimental trial. Images were captured from above using the setup depicted in Figure \ref{fig:schem}(c). Position tracking data (\S\ref{sec:methodposition}) was used to shift images into the traveling frame of the SurferBot and velocity fields are computed in a frame of reference moving with the SurferBot. Particle image velocimetry (PIV) analysis was performed in PIVlab \cite{thielicke2014pivlab}. Images were masked over the region occupied by the SurferBot and preprocessed using contrast limited adaptive histogram equalization (CLAHE) with 64 pixel window size and highpass filtering with 4 pixel kernel size. PIV was performed using a 64, 32, and 16 pixel interrogation window with 50 percent overlap for the first, second, and third pass, respectively. The velocity fields were smoothed and averaged over 54 consecutive frames. The velocity magnitude field (discussed in \S\ref{sec:resultssurface}) is presented at full resolution and overlaying velocity vectors were decimated by a factor of 5 in each direction.

\section{Results}
\label{sec:resultsAll}
When gently placed on the water surface, the SurferBot quickly generates an extended capillary wavefield (Figure \ref{speeds}(a)) and moves along its long axis in the direction of the battery (opposite the motor). The overall straightness of the trajectory was sensitive to the quality of the 3D print as well as fine spatial details of the pinning of the contact line.  Using six independent SurferBots and motors, 31 trajectories were collected as described in \S\ref{sec:methodposition}. The mean speed for the SurferBot was found to be $1.8\pm 0.3$ cm/s  (Figure \ref{speeds}(b)), or approximately 0.4 body lengths per second.  {\color{black} For comparison, the honeybee was measured to swim at $3.1\pm 0.8$ cm/s \cite{roh2019honeybees}, similar in magnitude to the SurferBot.} A sample trajectory is shown in Figure \ref{speeds}(c) with images of the SurferBot superimposed. Accompanying videos of the SurferBot and fluid motion are provided in the Supplementary Material.

%Describe general findings here (which direction they move, what speed, straightness comments, etc).

%Figure: Speed measurement histogram, overview photo of a straight trajectory, and aesthetic wave image.

\subsection{SurferBot kinematics}\label{resultskinematics}
\begin{figure}
\centering
\includegraphics{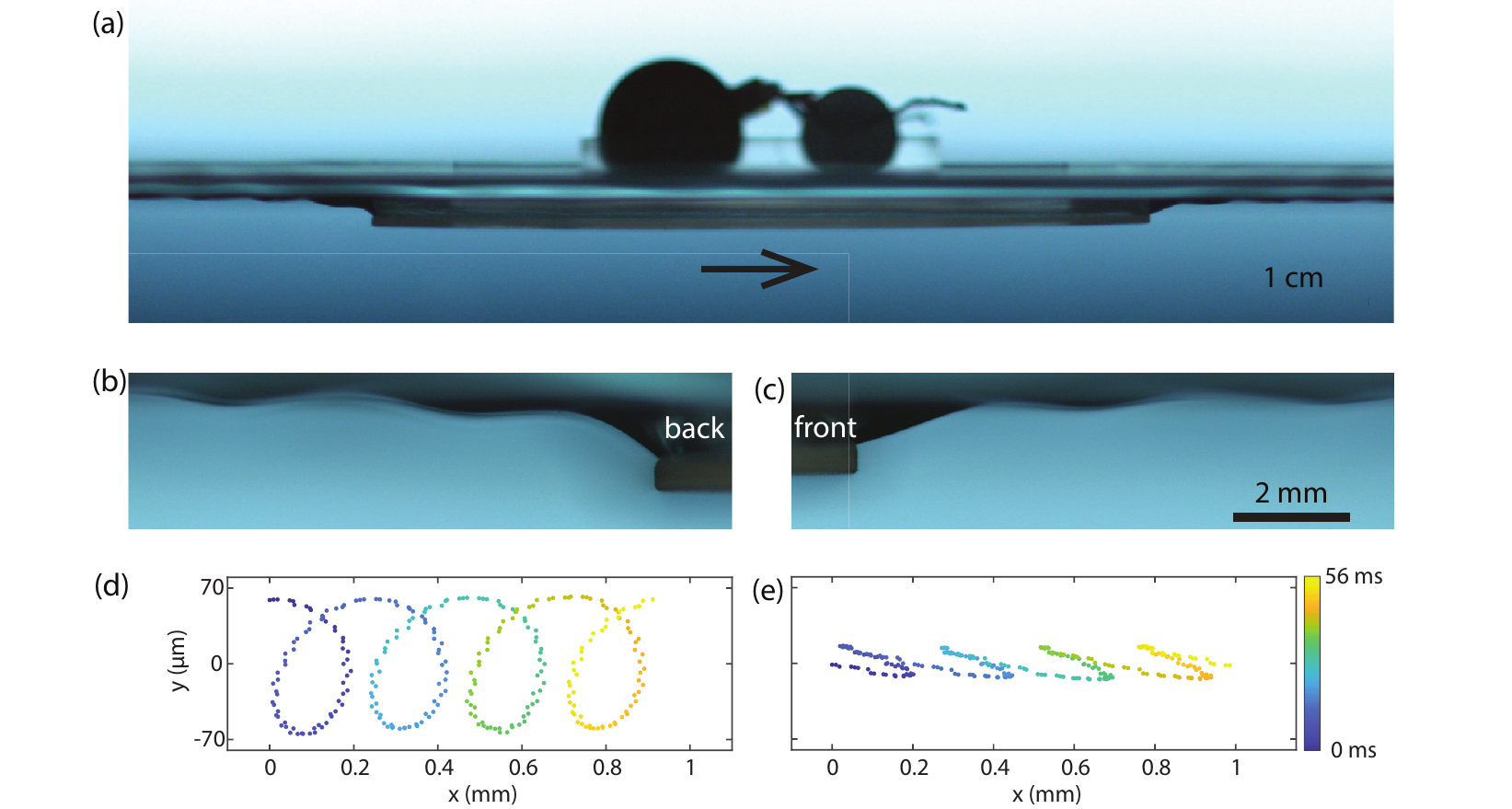}
\caption{(a) Still image from video taken from the side of a SurferBot moving from left to right. Higher magnification images of back (b) and front (c) edges of SurferBot from video were used for position tracking. Position of back (d) and front (e) edges shown over four motor cycles. Vertical positions are centered about their mean over the four cycles. See Supplementary Materials for related videos.}
\label{sidekinematics}
\end{figure}

The positions of the front and rear edges of the SurferBot were measured using the methods described in \S\ref{methodsideview}, with the results for four motor cycles presented in Figure \ref{sidekinematics}. Typically, the SurferBot would move about 400 $\mu$m forward and 200 $\mu$m backward in a full motor cycle, for a net forward motion of 200 $\mu$m per cycle. The amplitude of the vertical motion of the rear edge of the Surferbot was found to be $\sim 125$ $\mu$m (Figures \ref{sidekinematics}(b) and (d)), much greater than the amplitude at the front edge $\sim 25$ $\mu$m (Figures \ref{sidekinematics}(c) and (e)). As discussed in \S\ref{resultswavefield}, the waves propagating from the rear edge were also observed to have higher amplitude than in the front. The precise front-rear asymmetry of the amplitude of vertical motion of the SurferBot is likely due to the rear bias in positioning of the motor relative to the SurferBot center. The overall mass asymmetry also results in a larger static depression of the free surface from the equilibrium height at the rear edge than the front edge.

\subsection{Wavefield measurements}\label{resultswavefield}

\begin{figure}
\centering
\includegraphics{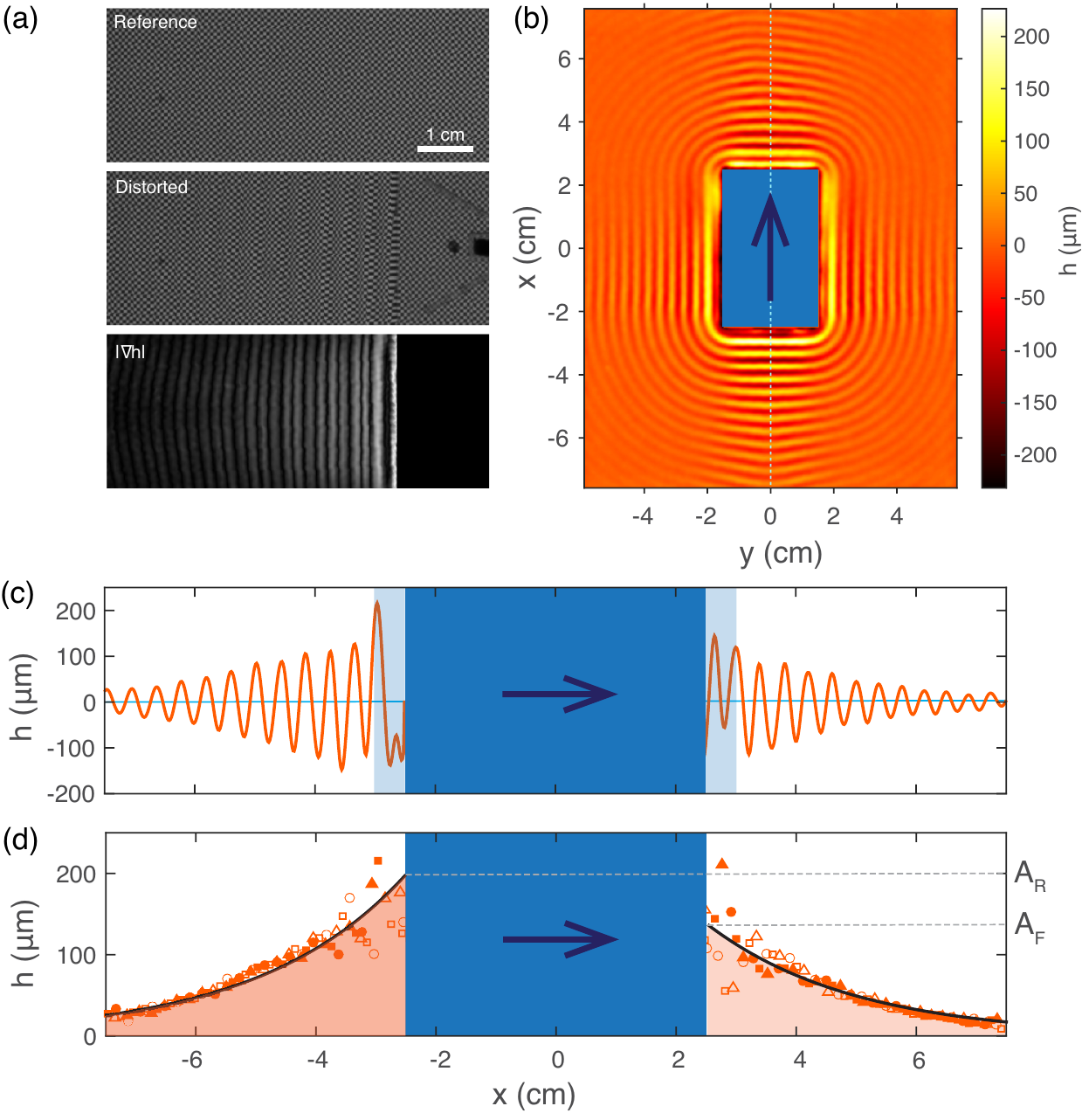}
\caption{Capillary wavefield of SurferBot. (a) Three images showcasing the FCD imaging technique for surface height reconstruction: a reference checkerboard is distorted by a wavefield from which the surface gradient $\nabla h$ can be deduced. (b) Reconstructed wavefield of a typical SurferBot. The arrow indicates the direction of travel of the SurferBot, while the dashed line denotes the centerline. (c) Wave height profile along centerline of SurferBot. The dark blue region (middle) represents the length of the SurferBot; distortion of the wavefield measurements occurs in the shaded blue regions at the front and back of the SurferBot. (d) Absolute value of local extrema in centerline wave profile superposed from many frames, to which a decaying exponential profile is fitted.  Filled/open symbols represent maxima/minima, with different marker shapes corresponding to different frames.}
\label{fig_waves}
\end{figure}

The wavefield for a typical SurferBot was measured using the methods described in \S\ref{fcd}, with the results presented in Figure \ref{fig_waves}. Figure \ref{fig_waves}(a) outlines the key ingredients of the FCD technique: an undistorted checkerboard reference pattern is distorted as the surfer passes through the field of view. These two images are then used to extract the surface height gradient $\nabla h(x,y)$, the relative magnitude of which is shown for the sample images in the bottom panel of Figure \ref{fig_waves}(a). In Figure \ref{fig_waves}(b), the complete reconstructed wavefield $h(x,y)$ of a SurferBot is presented. Typical wave amplitudes are seen to be on the order of 100 $\mu$m, consistent with the magnitude of the vertical motion of the SurferBot described in \S\ref{resultskinematics}.  

The measured wavefield along the centerline of the SurferBot is shown in Figure \ref{fig_waves}(c).  This curve is constructed from the data in Figure \ref{fig_waves}(b) by averaging the wave height over a strip of width 1 mm centered on the SurferBot and aligned with its long axis. From this data, it can be seen that the wave amplitude behind the SurferBot is slightly larger than in front, again consistent with the observations discussed in \S\ref{resultskinematics}.  The wave amplitude decays continuously away from the boat due to the effects of fluid viscosity as well as radial spreading associated with the finite size of the wave source.  The maxima and minima in the wave height from several frames at different oscillation phases are extracted and overlaid in Figure \ref{fig_waves}(d).  In order to quantify the decay rate and peak amplitude of the waves in the front and the rear, a simple decaying exponential function of form $h(x,0)=A_0 \exp(-\alpha |x-x_0|)$, where $x_0=\pm2.5$ cm represents the position of the front and rear edges of the SurferBot, is fit to the amplitude data using a least-squares method. Only data for $|x-x_0|>0.5$ cm was used to determine the exponential fit parameters, as the data for $|x-x_0|<0.5$ cm was distorted as discussed in \S\ref{fcd}.  The data for the front and rear of the SurferBot was well-approximated by a spatial decay rate $\alpha=0.41 \pm 0.04$ cm$^{-1}$. Using this value of $\alpha$, the peak wave amplitude at the boat edge was found to be approximately $A_0 = A_R=199 \pm 10 \ \mu$m at the rear and $A_0 = A_F=138 \pm 10 \ \mu$m in the front. The fitted curves are shown as the solid lines in Figure \ref{fig_waves}(d).  Finally, by extracting the spacing between successive peaks of the wavefield, the wavelength of the capillary wavefield was found to be $\lambda = 3.97 \pm 0.20$ mm.  This wavelength is consistent with the dispersion relation for deep-water capillary waves {\color{black} \cite{de2004capillarity}}
\begin{equation}
    \omega^2 = \frac{\sigma}{\rho}k^3,
\end{equation}
where $\sigma$ is the surface tension, $k = 2\pi/\lambda$ is the wavenumber, $\omega = 2 \pi f$, and $f$ is taken to be the typical oscillation frequency range (in Hz) of the vibration motors.

The difference in wave amplitude between the front and rear represents an unbalanced momentum flux, or radiation stress, and thus a mechanism for propulsion. This mechanism of propulsion due to fore-aft asymmetry of the wavefield is encountered in both the swimming honeybee \cite{roh2019honeybees} and in recently discovered synthetic surfers driven to self-propel by global vibration of a fluid bath \cite{ho2021capillary}. 

Classical results for wave momentum flux of capillary waves predict the resultant force (per unit length) associated with an outwardly propagating plane wave as $3\sigma k^2 A^2/4$ where $A$ is the wave amplitude \cite{longuet1964radiation}. Thus for the SurferBot, we can estimate the wave-induced net propulsive force as 
\begin{equation}
    F_W=\frac{3}{4}\sigma k^2 (A_R^2-A_F^2)w,
\end{equation}
where $w=3$ cm is the width of the base.  From our estimates of the wave amplitudes $A_R$ and $A_F$ and the fluid properties of water, we find $F_W$ = 8.3 dynes (83 $\mu N$). For comparison, the net wave thrust produced for the surface-bound honeybee was estimated to be 5 dynes (50 $\mu N$) \cite{roh2019honeybees}, of very similar magnitude to the SurferBot. Of the force estimates provided for the honeybee, the asymmetric wave momentum flux represented the largest contribution to the total thrust. 

\subsection{Surface flow measurements}\label{sec:resultssurface}
\begin{figure}
\centering
\includegraphics{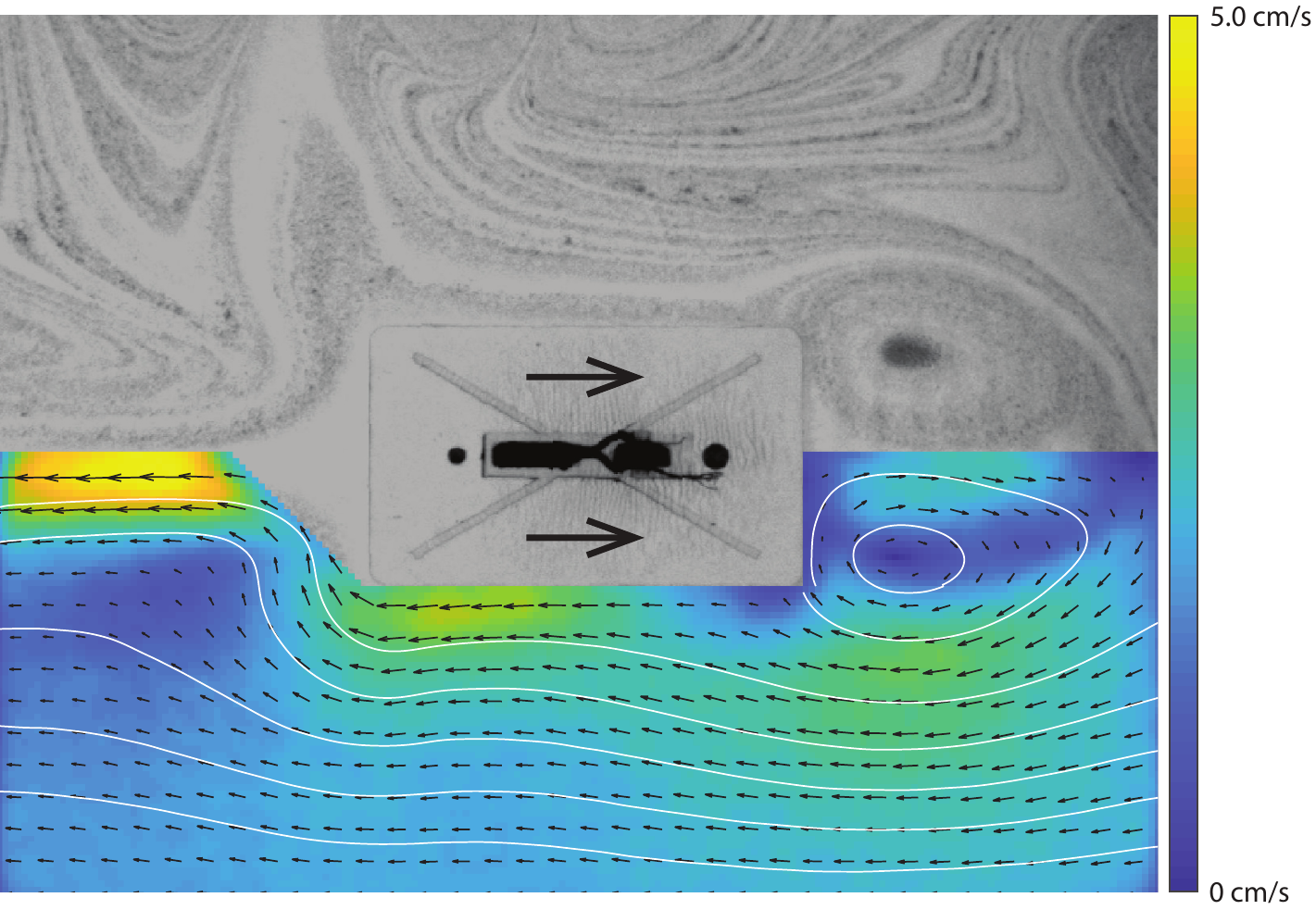}
\caption{Raw image from video used for surface velocity measurements (upper half) and resulting velocity field (bottom half). Results pictured are averaged over 54 frames and presented in the frame of the SurferBot (moving left to right). Presented data include velocity magnitude (background pseudocolor plot), velocity field (black vectors), and streamlines (white contour lines).  See Supplementary Materials for video of particle motion.}
\label{surfaceflow}
\end{figure}
The velocity field on the free surface, measured using PIV (\S\ref{methodsurface}), is presented in Figure \ref{surfaceflow}. The velocity is presented in the frame of the SurferBot which is moving left to right. We observe that an outward jet emerges at the front and rear of the SurferBot, the front jet slowing to a stagnation point and returning to form a bound counter-rotating vortex pair.  The peak speed of the fluid is faster in the rear, consistent with the larger wave amplitude in that region. Although the buoyant hollow glass microspheres used as tracing particles are stirred before each trial, they develop streamline-like striations as the flow persists, pictured in the upper half of Figure \ref{surfaceflow}. This is especially evident in the front vortex pair, whose closed streamlines trap clumps of particles. Results near the rear edge of the SurferBot were excluded due to low seeding density. A control volume estimate of the momentum imparted by the SurferBot with a depth length scale of $(2k)^{-1}$ \cite{stokes1847theory} yields forces less than $0.1$ dyne, however more precise quantitative estimates of momentum flux are difficult without velocity measurements in the bulk.

While Stokes drift is perhaps the most familiar mechanism for wave-induced fluid flows \cite{stokes1847theory}, other related mechanisms occuring due to nonlinear flow interactions are possible \cite{longuet1953mass, filatov2016nonlinear}.  Such flows represent an active area of ongoing fundamental research \cite{monismith2020stokes,parfenyev2020large}. The focused jets emerging from the front and rear of the SurferBot are similar in form to prior observations of steady surface flows from finite planar wavemakers \cite{punzmann2014generation} and likely share similar fluid-mechanical origins.   

%\subsection{Drag and thrust estimates}
%Dan.

\subsection{Other modes of locomotion}
%Figure 7: 4 photos depicting each state (regular, upside down, fully submerged, dry land).  Perhaps one where it walks out of water onto land (or vice versa) via a ramp? 

\begin{figure}
\centering
\includegraphics{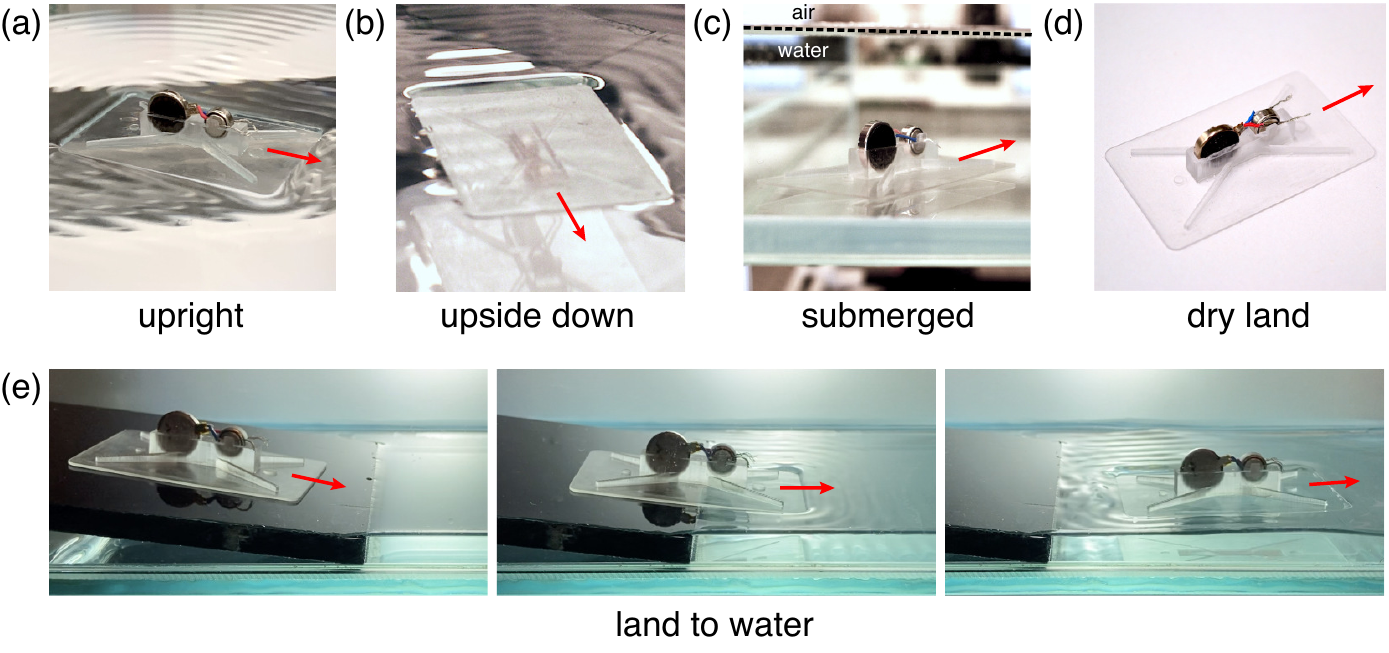}
\caption{(a) SurferBot upright at the interface. (b) SurferBot inverted at the interface. (c) SurferBot fully submerged beneath the water. (d) SurferBot on dry land on a sheet of paper.  (e) SurferBot transitioning from motion on dry land (acrylic) to the water surface. See Supplementary Materials for related video.}
\label{transp}
\end{figure}

In various separate experimental trials, it was observed that the SurferBot can also propel in a variety of configurations and on substrates other than water, each of which are highlighted in Figure \ref{transp}.  In particular, a SurferBot that is placed either upright (Figure \ref{transp}(a)), inverted (Figure \ref{transp}(b)), or fully submerged (Figure \ref{transp}(c)) will all move forward in the direction of the battery. A SurferBot that is placed on a flat surface (Figure \ref{transp}(d)) typically also moves in the direction towards the battery (albeit in a somewhat more erratic fashion), but can also be observed to move towards the motor, depending on the substrate properties and flatness of the SurferBot base. This {\color{black} erratic motion} is {\color{black} somewhat} akin to \emph{BristleBot} dynamics (discussed in \S\ref{sec:discussion}) \cite{giomi2013swarming, becker2014mechanics, cicconofri2015motility}. Finally, it is possible for the SurferBot to transition between these various modes of locomotion. For instance, Figure \ref{transp}(e) showcases an event where the SurferBot embarks on an interfacial voyage after first traveling down a dry acrylic ramp. The amphibious nature of the SurferBot was unintended, but may prove beneficial for robustness and utility in applications \cite{dudek2007aqua}. {\color{black} The present work has focused on documenting the propulsion of the SurferBot in the intended upright configuration; additional work is needed to fully characterize the alternative modes of locomotion briefly discussed in this section and depicted in Figure \ref{transp}.}

\section{Discussion}
\label{sec:discussion}
%This should discuss the significance of the results and compare them with previous work using relevant references.

% our estimates of the propulsion force suggests that unbalanced radiation pressure associated with the asymmetric wavefield is the primary driver for propelling the SurferBot.  Furthermore, we anticipated that the bot is principally resisted by skin friction along its base.  Estimating the flow beneath the bot as a linear shear flow, allows us to estimate the viscous resistance force

% \begin{equation}
%     F_D = \mu \frac{U}{h}wl
% \end{equation}
% where is $\mu$ is the viscosity of water, and $h$ is the depth of fluid directly beneath the bot (estimated at 3 mm).  

%Drag estimate?

The SurferBot is a low-cost, simple robotic device consisting of a flat plate supporting an off-centered vibration motor. As a result of the detailed measurements described in \S \ref{resultskinematics}--\S\ref{sec:resultssurface}, a consistent physical picture describing the propulsion of the SurferBot emerges. The off-centered placement of the vibration motor leads to amplified vertical motion of the rear of the SurferBot, resulting in a fore-aft asymmetric capillary wavefield.  These capillary waves carry with them excess momentum, with the asymmetry resulting in a net forward thrust force on the SurferBot. The generation and propagation of the capillary wavefields also gave rise to rich surface flow patterns, primarily defined by a pronounced jet at the rear of the SurferBot and a trapped counter-rotating vortex pair at the front. While considerably simpler geometrically, the asymmetric wavefield and steady surface flow pattern share remarkable similarities with the hydrofoiling honeybee \cite{roh2019honeybees}.

Among other robotic devices developed to operate at an air-water interface, there exist several other prior designs that leverage on-board vibration sources for self-propulsion \cite{suzuki2007water, becker2013amphibious, lee2019milli, cocuzza2021vibration}.  {\color{black} These devices move at similar speeds to the SurferBot, in the range of 1-10 cm/s depending on the design details and operational parameters.}  In contrast to the rigid-body response of the SurferBot, each of these prior designs relies on excitation of resonant modes of a flexible robotic structure to achieve kinematics inspired by water striders.  Furthermore, the potential role of capillary waves in the propulsion mechanism of such vibrobots has been suggested \cite{lee2019milli}, but not yet clearly demonstrated.

Another class of interfacial robots with similar hydrodynamics to the SurferBot are centimeter-scale robots propelled \emph{via} electrowetting \cite{chung2009electrowetting,yuan2015mechanism, jiang2020triboelectric}.  For these devices, one of the robot's sides was covered with an electrode through which an AC voltage could be applied.  This oscillating potential results in oscillations of the contact angle and thus the contact line position, generating an outward propagating capillary wave and steady surface flow.  While a creative solution that involves no moving parts, this device does require relatively high voltages (around 100 V or greater) and more complicated electronics for propulsion to be realized. 

A category of popular simple robots that shares {\color{black} some} similarities with the SurferBot is the BristleBot \cite{giomi2013swarming, becker2014mechanics, cicconofri2015motility}.  In its simplest form, the BristleBot has the same electronics package as the SurferBot, typically consisting of only a vibration motor and coin cell battery.  Rather than being mounted to a flat plate, however, the vibration motor is mounted to the top of a surface with angled bristles, such as the head of the toothbrush, and operates on dry land.  Due to their relative simplicity, safety, and low-cost, BristleBots frequently appear in pedagogical and outreach applications \cite{ross2017phototropic}.  We anticipate that the SurferBot, the BristleBot's seafaring cousin, could also be applied in educational settings, as well as being used to study novel modes of collective dynamics at the fluid interface.

Although more sophisticated manufacturing techniques were used to develop the final SurferBot (see \S\ref{3d}), a decidedly simpler construction can be used for pedagogical purposes using inexpensive and readily available materials. For example, the 3D-printed SurferBot base may be replaced by an appropriately cut thin piece of plastic, such as an overhead transparency film. The arm-like supports that improve rigidity are not essential for motion, but could be replaced with matchsticks or toothpicks, while the motor and battery can be secured with double-sided tape or modeling clay. While the straightness of the SurferBot trajectory may suffer as a result of this more rudimentary design, the construction and deployment of the SurferBot should make for an engaging classroom activity and striking demonstration of fundamental fluid mechanics in action.

Finally, in addition to those discussed previously, other small-scale vibrobots have been shown to propel along or within other media, such as along a granular bed \cite{quillen2019light} or freely immersed in a fluid \cite{quillen2016coin}.  At vastly larger length-scales, forced rocking motion of a canoe has also been shown to lead to self-propulsion \emph{via} gravity wave effects, a technique employed in competitive ``gunwale bobbing'' \cite{benham2022gunwale}.

\section{Conclusion}
\label{sec:conclusion}
%This section should be used to highlight the novelty and significance of the work, and any plans for future relevant work.

In this work, we have developed a simple, low-cost vibrating robot capable of self-propelling along an air-water interface. The fundamental fluid mechanics underpinning self-propulsion share striking similarities to honeybees trapped at the air-water interface. Steady rectilinear motion is achieved through an unbalanced momentum transfer to the fluid interface as a result of asymmetric vibration.  

Future work will include exploring the vast parameter space exposed by the present study. Preliminary observations suggest that the speed of the SurferBot is particularly sensitive to motor placement and SurferBot length.  {\color{black} In particular, placing the motor closer to rear of the SurferBot increases its overall speed, but with the trade-off that it appears more prone to rotation.}  Furthermore, it would be interesting to study the dynamics of the SurferBot in confinement \cite{bechinger2016active}: due to the long-range nature of the waves and flows, interesting dynamics are anticipated when the SurferBots are forced to interact with boundaries or other SurferBots. For example, SurferBots confined to quasi-one-dimensional channels may exhibit collective self-sustaining waves analogous to those observed in traffic jams \cite{flynn2009self} and more recently in lattices of confined levitating droplets \cite{thomson2020collective}. The SurferBot may also find fundamental applications as a macroscopic entity with both wave and particle-like features \cite{bush2020hydrodynamic,schiebel2019mechanical}, or as highly tunable constituents in studies of collective motion of surface swimmers \cite{klotsa2019above,feldmann2021can,ho2021capillary,devereux2021whirligig}. 

With a view toward real-life applications, other energy sources that allow for longer experimental run time should be explored, such as solar power \cite{fortunic2017design}, microbial fuel cells \cite{philamore2015row}, or other chemical means \cite{wang2019floating}.  Lastly, the current design uses only a single motor at a fixed frequency, and thus only moves along a linear trajectory. The addition of a second (controllable) vibration motor may enable fully two-dimensional navigation of the SurferBot as it enters uncharted waters.

\section{Acknowledgements}

We gratefully acknowledge the financial support of the Office of Naval Research (ONR N00014-21-1-2816 and ONR N00014-21-1-2670) and Brown University (Karen T. Romer Undergraduate Research Award).  We also thank Luke Alventosa and Jack-William Barotta for advice during development of the SurferBot and useful feedback on the manuscript, and Davis Smith for additional support and feedback on the aesthetics.

\newpage

\bibliographystyle{unsrt}
\bibliography{surferbib.bib}

\begin{thebibliography}{10}

\bibitem{bush2006walking}
J.~W.~M. Bush and D.~L. Hu.
\newblock Walking on water: biolocomotion at the interface.
\newblock {\em Annual Review of Fluid Mechanics}, 38:339--369, 2006.

\bibitem{vella2015floating}
D.~Vella.
\newblock Floating versus sinking.
\newblock {\em Annual Review of Fluid Mechanics}, 47:115--135, 2015.

\bibitem{hu2003hydrodynamics}
D.~L. Hu, B.~Chan, and J.~W.~M. Bush.
\newblock The hydrodynamics of water strider locomotion.
\newblock {\em Nature}, 424(6949):663--666, 2003.

\bibitem{voise2010management}
J.~Voise and J.~Casas.
\newblock The management of fluid and wave resistances by whirligig beetles.
\newblock {\em Journal of the Royal Society Interface}, 7(43):343--352, 2010.

\bibitem{ngo2014hydrodynamics}
V.~Ngo and M.~J. Mc{H}enry.
\newblock The hydrodynamics of swimming at intermediate {R}eynolds numbers in
  the water boatman ({C}orixidae).
\newblock {\em Journal of Experimental Biology}, 217(15):2740--2751, 2014.

\bibitem{buhler2007impulsive}
O.~B{\"u}hler.
\newblock Impulsive fluid forcing and water strider locomotion.
\newblock {\em Journal of Fluid Mechanics}, 573:211--236, 2007.

\bibitem{gao2011numerical}
P.~Gao and J.~Feng.
\newblock A numerical investigation of the propulsion of water walkers.
\newblock {\em Journal of Fluid Mechanics}, 668:363--383, 2011.

\bibitem{steinmann2018unsteady}
T.~Steinmann, M.~Arutkin, P.~Cochard, E.~Rapha{\"e}l, J.~Casas, and
  M.~Benzaquen.
\newblock Unsteady wave pattern generation by water striders.
\newblock {\em Journal of Fluid Mechanics}, 848:370--387, 2018.

\bibitem{hu2010water}
D.~L. Hu, M.~Prakash, B.~Chan, and J.~W.~M. Bush.
\newblock Water-walking devices.
\newblock In {\em Animal Locomotion}, pages 131--140. Springer, 2010.

\bibitem{yuan2012bio}
J.~Yuan and S.~K. Cho.
\newblock Bio-inspired micro/mini propulsion at air-water interface: {A}
  review.
\newblock {\em Journal of Mechanical Science and Technology},
  26(12):3761--3768, 2012.

\bibitem{burton2013biomimicry}
L.~J. Burton, N.~Cheng, C.~Vega, J.~Andr{\'e}s, and J.~W.~M. Bush.
\newblock Biomimicry and the culinary arts.
\newblock {\em Bioinspiration \& Biomimetics}, 8(4):044003, 2013.

\bibitem{ozcan2014stride}
O.~Ozcan, H.~Wang, J.~D. Taylor, and M.~Sitti.
\newblock Stride {II}: a water strider-inspired miniature robot with circular
  footpads.
\newblock {\em International Journal of Advanced Robotic Systems}, 11(6):85,
  2014.

\bibitem{koh2015jumping}
J-S. Koh, E.~Yang, G-P. Jung, S-P. Jung, J.~H. Son, S-I. Lee, P.G. Jablonski,
  R.~J. Wood, H-Y. Kim, and K-J. Cho.
\newblock Jumping on water: {S}urface tension--dominated jumping of water
  striders and robotic insects.
\newblock {\em Science}, 349(6247):517--521, 2015.

\bibitem{philamore2015row}
H.~Philamore, J.~Rossiter, A.~Stinchcombe, and I.~Ieropoulos.
\newblock Row-bot: {A}n energetically autonomous artificial water boatman.
\newblock In {\em 2015 IEEE/RSJ International Conference on Intelligent Robots
  and Systems (IROS)}, pages 3888--3893. IEEE, 2015.

\bibitem{chen2018controllable}
Y.~Chen, N.~Doshi, B.~Goldberg, H.~Wang, and R.~J. Wood.
\newblock Controllable water surface to underwater transition through
  electrowetting in a hybrid terrestrial-aquatic microrobot.
\newblock {\em Nature Communications}, 9(1):1--11, 2018.

\bibitem{grosjean2018surface}
G.~Grosjean, M.~Hubert, Y.~Collard, S.~Pillitteri, and N.~Vandewalle.
\newblock Surface swimmers, harnessing the interface to self-propel.
\newblock {\em The European Physical Journal E}, 41(11):1--10, 2018.

\bibitem{feldmann2021can}
D.~Feldmann, R.~Das, and B.-E. Pinchasik.
\newblock How can interfacial phenomena in nature inspire smaller robots.
\newblock {\em Advanced Materials Interfaces}, 8(1):2001300, 2021.

\bibitem{timm2021remotely}
M.~L. Timm, S.~J. Kang, J.~P. Rothstein, and H.~Masoud.
\newblock A remotely controlled {M}arangoni surfer.
\newblock {\em Bioinspiration \& Biomimetics}, 16(6):066014, 2021.

\bibitem{roh2019honeybees}
C.~Roh and M.~Gharib.
\newblock Honeybees use their wings for water surface locomotion.
\newblock {\em Proceedings of the National Academy of Sciences},
  116(49):24446--24451, 2019.

\bibitem{pan2009miniature}
{Q. Pan and M. Wang}.
\newblock {Miniature boats with striking loading capacity fabricated from
  superhydrophobic copper meshes}.
\newblock {\em {ACS Applied Materials \& Interfaces}}, {1}({2}):{420--423},
  {2009}.

\bibitem{wildeman2018real}
S.~Wildeman.
\newblock Real-time quantitative {S}chlieren imaging by fast {F}ourier
  demodulation of a checkered backdrop.
\newblock {\em Experiments in Fluids}, 59(6):1--13, 2018.

\bibitem{moisy2009synthetic}
F.~Moisy, M.~Rabaud, and K.~Salsac.
\newblock A synthetic {S}chlieren method for the measurement of the topography
  of a liquid interface.
\newblock {\em Experiments in Fluids}, 46(6):1021--1036, 2009.

\bibitem{grediac2016grid}
M.~Grediac, F.~Sur, and B.~Blaysat.
\newblock The grid method for in-plane displacement and strain measurement: {A}
  review and analysis.
\newblock {\em Strain}, 52(3):205--243, 2016.

\bibitem{wildemangithub}
S.~Wildeman.
\newblock {MATLAB} implementations of {FCD} and {DIC+OF} methods.
\newblock \url{https://github.com/swildeman/fcd}, 2017.

\bibitem{d2009inverse}
J.~D{’}Errico.
\newblock Inverse (integrated) gradient.
\newblock
  \url{https://www.mathworks.com/matlabcentral/fileexchange/9734-inverse-integrated-gradient},
  2013.

\bibitem{damiano2016surface}
A.~P. Damiano, P-T. Brun, D.~M. Harris, C.~A. Galeano-Rios, and J.~W.~M. Bush.
\newblock Surface topography measurements of the bouncing droplet experiment.
\newblock {\em Experiments in Fluids}, 57(10):1--7, 2016.

\bibitem{harris2017visualization}
D.~M. Harris, J.~Quintela, V.~Prost, P-T. Brun, and J.~W.~M. Bush.
\newblock Visualization of hydrodynamic pilot-wave phenomena.
\newblock {\em Journal of Visualization}, 20(1):13--15, 2017.

\bibitem{colombi2021three}
R.~Colombi, M.~Schl{\"u}ter, and A.~Kameke.
\newblock Three-dimensional flows beneath a thin layer of 2{D} turbulence
  induced by {F}araday waves.
\newblock {\em Experiments in Fluids}, 62(1):1--13, 2021.

\bibitem{thielicke2014pivlab}
W.~Thielicke and E.~Stamhuis.
\newblock {PIV}lab--towards user-friendly, affordable and accurate digital
  particle image velocimetry in {MATLAB}.
\newblock {\em Journal of Open Research Software}, 2(1), 2014.

\bibitem{de2004capillarity}
{P.-G. De Gennes and F. Brochard-Wyart and D. Qu{\'e}r{\'e}}.
\newblock {\em {Capillarity and wetting phenomena: drops, bubbles, pearls,
  waves}}, volume {315}.
\newblock {Springer}, {2004}.

\bibitem{ho2021capillary}
I.~Ho, G.~Pucci, A.~U. Oza, and D.~M. Harris.
\newblock Capillary surfers: wave-driven particles at a fluid interface.
\newblock {\em arXiv preprint arXiv:2102.11694}, 2021.

\bibitem{longuet1964radiation}
M.~S. Longuet-Higgins and R.~W. Stewart.
\newblock Radiation stresses in water waves; a physical discussion, with
  applications.
\newblock In {\em Deep Sea Research and Oceanographic Abstracts}, volume~11,
  pages 529--562. Elsevier, 1964.

\bibitem{stokes1847theory}
G.~G. Stokes.
\newblock On the theory of oscillatory waves.
\newblock {\em Transactions of the Cambridge Philosophical Society}, 1847.

\bibitem{longuet1953mass}
M.~S. Longuet-Higgins.
\newblock Mass transport in water waves.
\newblock {\em Philosophical Transactions of the Royal Society of London.
  Series A, Mathematical and Physical Sciences}, 245(903):535--581, 1953.

\bibitem{filatov2016nonlinear}
S.~V. Filatov, V.~M. Parfenyev, S.~S. Vergeles, M.~Y. Brazhnikov, A.~A.
  Levchenko, and V.~V. Lebedev.
\newblock Nonlinear generation of vorticity by surface waves.
\newblock {\em Physical Review Letters}, 116(5):054501, 2016.

\bibitem{monismith2020stokes}
S.~G. Monismith.
\newblock Stokes drift: theory and experiments.
\newblock {\em Journal of Fluid Mechanics}, 884, 2020.

\bibitem{parfenyev2020large}
V.~M. Parfenyev and S.~S. Vergeles.
\newblock Large-scale vertical vorticity generated by two crossing surface
  waves.
\newblock {\em Physical Review Fluids}, 5(9):094702, 2020.

\bibitem{punzmann2014generation}
H.~Punzmann, N.~Francois, H.~Xia, G.~Falkovich, and M.~Shats.
\newblock Generation and reversal of surface flows by propagating waves.
\newblock {\em Nature Physics}, 10(9):658--663, 2014.

\bibitem{giomi2013swarming}
L.~Giomi, N.~Hawley-Weld, and L.~Mahadevan.
\newblock Swarming, swirling and stasis in sequestered bristle-bots.
\newblock {\em Proceedings of the Royal Society A: Mathematical, Physical and
  Engineering Sciences}, 469(2151):20120637, 2013.

\bibitem{becker2014mechanics}
F.~Becker, S.~Boerner, V.~Lysenko, I.~Zeidis, and K.~Zimmermann.
\newblock On the mechanics of bristle-bots-modeling, simulation and
  experiments.
\newblock In {\em ISR/Robotik 2014; 41st international symposium on robotics},
  pages 1--6. VDE, 2014.

\bibitem{cicconofri2015motility}
G.~Cicconofri and A.~DeSimone.
\newblock Motility of a model bristle-bot: A theoretical analysis.
\newblock {\em International Journal of Non-Linear Mechanics}, 76:233--239,
  2015.

\bibitem{dudek2007aqua}
G.~Dudek, P.~Giguere, C.~Prahacs, S.~Saunderson, J.~Sattar, L-A. Torres-Mendez,
  M.~Jenkin, A.~German, A.~Hogue, and A.~et~al. Ripsman.
\newblock Aqua: {A}n amphibious autonomous robot.
\newblock {\em Computer}, 40(1):46--53, 2007.

\bibitem{suzuki2007water}
K.~Suzuki, H.~Takanobu, K.~Noya, H.~Koike, and H.~Miura.
\newblock Water strider robots with microfabricated hydrophobic legs.
\newblock In {\em 2007 IEEE/RSJ International Conference on Intelligent Robots
  and Systems}, pages 590--595. IEEE, 2007.

\bibitem{becker2013amphibious}
F.~Becker, K.~Zimmermann, T.~Volkova, and V.T. Minchenya.
\newblock An amphibious vibration-driven microrobot with a piezoelectric
  actuator.
\newblock {\em Regular and Chaotic Dynamics}, 18(1):63--74, 2013.

\bibitem{lee2019milli}
K.~Y. Lee, L.~Wang, J.~Qu, and K.~R. Oldham.
\newblock Milli-scale biped vibratory water strider.
\newblock In {\em 2019 International Conference on Manipulation, Automation and
  Robotics at Small Scales (MARSS)}, pages 1--6. IEEE, 2019.

\bibitem{cocuzza2021vibration}
S.~Cocuzza, A.~Doria, and M.~Reis.
\newblock Vibration-based locomotion of an amphibious robot.
\newblock {\em Applied Sciences}, 11(5):2212, 2021.

\bibitem{chung2009electrowetting}
S.~K. Chung, K.~Ryu, and S.~K. Cho.
\newblock Electrowetting propulsion of water-floating objects.
\newblock {\em Applied Physics Letters}, 95(1):014107, 2009.

\bibitem{yuan2015mechanism}
J.~Yuan and S.~K. Cho.
\newblock Mechanism and flow measurement of ac electrowetting propulsion on
  free surface.
\newblock {\em Experiments in Fluids}, 56(3):1--10, 2015.

\bibitem{jiang2020triboelectric}
D.~Jiang, Z.~Fan, H.~Wang, M.~Xu, G.~Chen, Y.~Song, and Z.~L. Wang.
\newblock Triboelectric nanogenerator powered electrowetting-on-dielectric
  actuator for concealed aquatic microbots.
\newblock {\em ACS Nano}, 14(11):15394--15402, 2020.

\bibitem{ross2017phototropic}
R.~Ross, J.~Stanger, and A.~Console.
\newblock Phototropic {B}ristle{B}ot activity for robotics and {STEM}
  engagement.
\newblock In {\em 2017 IEEE International Conference on Mechatronics (ICM)},
  pages 425--430. IEEE, 2017.

\bibitem{quillen2019light}
A.~C. Quillen, R.~C. Nelson, H.~Askari, K.~Chotkowski, E.~Wright, and J.~K.
  Shang.
\newblock A light-weight vibrational motor powered recoil robot that hops
  rapidly across granular media.
\newblock {\em Journal of Mechanisms and Robotics}, 11(6), 2019.

\bibitem{quillen2016coin}
A.~C. Quillen, H.~Askari, D.~H. Kelley, T.~Friedmann, and P.~W. Oakes.
\newblock A coin vibrational motor swimming at low {R}eynolds number.
\newblock {\em Regular and Chaotic Dynamics}, 21(7):902--917, 2016.

\bibitem{benham2022gunwale}
G.~P. Benham, O.~Devauchelle, S.~W. Morris, and J.~A. Neufeld.
\newblock Gunwale bobbing.
\newblock {\em arXiv preprint arXiv:2201.01533}, 2022.

\bibitem{bechinger2016active}
C.~Bechinger, R.~Di~Leonardo, H.~L{\"o}wen, C.~Reichhardt, G.~Volpe, and
  G.~Volpe.
\newblock Active particles in complex and crowded environments.
\newblock {\em Reviews of Modern Physics}, 88(4):045006, 2016.

\bibitem{flynn2009self}
M.~R. Flynn, A.~R. Kasimov, J-C. Nave, R.~R. Rosales, and B.~Seibold.
\newblock Self-sustained nonlinear waves in traffic flow.
\newblock {\em Physical Review E}, 79(5):056113, 2009.

\bibitem{thomson2020collective}
S.~J. Thomson, M.~M.~P. Couchman, and J.~W.~M. Bush.
\newblock Collective vibrations of confined levitating droplets.
\newblock {\em Physical Review Fluids}, 5(8):083601, 2020.

\bibitem{bush2020hydrodynamic}
J.~W.~M. Bush and A.~U. Oza.
\newblock Hydrodynamic quantum analogs.
\newblock {\em Reports on Progress in Physics}, 84(1):017001, 2020.

\bibitem{schiebel2019mechanical}
P.~E. Schiebel, J.~M. Rieser, A.~M. Hubbard, L.~Chen, D.~Z. Rocklin, and D.~I.
  Goldman.
\newblock Mechanical diffraction reveals the role of passive dynamics in a
  slithering snake.
\newblock {\em Proceedings of the National Academy of Sciences},
  116(11):4798--4803, 2019.

\bibitem{klotsa2019above}
D.~Klotsa.
\newblock As above, so below, and also in between: mesoscale active matter in
  fluids.
\newblock {\em Soft Matter}, 15(44):8946--8950, 2019.

\bibitem{devereux2021whirligig}
H.~L. Devereux, C.~R. Twomey, M.~S. Turner, and S.~Thutupalli.
\newblock Whirligig beetles as corralled active {B}rownian particles.
\newblock {\em Journal of the Royal Society Interface}, 18(177):20210114, 2021.

\bibitem{fortunic2017design}
J.~E.~P. Fortuni{\'c}.
\newblock {\em Design and implementation of a bristle-bot swarm system}.
\newblock Pontificia Universidad Catolica del Peru-CENTRUM Catolica (Peru),
  2017.

\bibitem{wang2019floating}
Y.~Wang, Y.~Jiang, H.~Wu, and Y.~Yang.
\newblock Floating robotic insects to obtain electric energy from water surface
  for realizing some self-powered functions.
\newblock {\em Nano Energy}, 63:103810, 2019.

\end{thebibliography}

\end{document}